\documentclass[
 reprint,
superscriptaddress,
 amsmath,amssymb,
 aps,
]{revtex4-2}
\usepackage{mathtools}
\usepackage{graphicx}
\usepackage{dcolumn}
\usepackage{bm}
\usepackage{siunitx}
\usepackage[utf8]{inputenc}
\usepackage[T1]{fontenc}

\makeatletter
\renewcommand*{\fnum@figure}{{\normalfont\bfseries \figurename~\thefigure}}
\renewcommand*{\@caption@fignum@sep}{\textbf{\usepackage{.} }}
\makeatother
\usepackage{amsmath}
\usepackage{setspace}
\usepackage {algpseudocode}
\usepackage{lipsum}
\usepackage{xurl}
\makeatletter
\renewcommand*{\fnum@figure}{{\normalfont\bfseries \figurename~\thefigure}}
\renewcommand*{\@caption@fignum@sep}{\textbf{. }}
\makeatother
\renewcommand{\thefigure}{\arabic{figure}}
\usepackage{hyperref}

\begin{document}

\preprint{APS/123-QED}
\raggedbottom
\title{Inverse Engineering of Optical Constants in Photochromic Micron-Scale Hybrid Films}

\author{Bahrem~Serhat~Danis}
\email{bdanis23@ku.edu.tr}
\affiliation{\mbox{Department of Electrical and Electronics Engineering, Koç University, Istanbul, 34450, Turkey}}

\author{Amin~Tabatabaei~Mohseni}
\affiliation{\mbox{Nano Science and Nano Engineering Department, Istanbul Technical University, Istanbul, 34469, Turkey}}

\author{Smagul~Karazhanov}
\affiliation{\mbox{Department of Solar Energy, Institute for Energy Technology (IFE), Kjeller, 2027, Norway}}
\affiliation{\mbox{Thin Film Laboratories, Institute of Solid State Physics, University of Latvia, LV-1063 Riga, Latvia}}

\author{Esra~Zayim}
\email[Corresponding author: ]{ozesra@itu.edu.tr}
\affiliation{\mbox{Physics Engineering Department, Istanbul Technical University, Istanbul, 34469, Turkey}}

\begin{abstract}
Photochromic materials enable dynamic optical modulation through reversible transitions between distinct absorption states, with broad potential for smart windows, adaptive optics, and reconfigurable photonic devices. Micron-scale photochromic hybrid films present a particularly attractive platform for these applications, combining straightforward preparation with substantial optical modulation and scalability for high-volume fabrication. However, rational design of such films remains fundamentally constrained by the absence of well-defined optical constants. Unlike homogeneous thin films, micron-scale hybrid photochromic materials comprise active particles dispersed non-uniformly within polymer matrices. Conventional first-principles electromagnetic simulations face substantial computational costs and discrepancies between simulated and experimental particle distributions. Here, we introduce a data-driven framework that extracts effective optical constants directly from minimal experimental transmittance measurements. Our dual-state effective model approximates the complex inhomogeneous photochromic layer as a compressed homogeneous medium characterized by pseudo-refractive indices and pseudo-extinction coefficients for both pristine and UV-irradiated states. Through systematic optimization against experimental data from tungsten oxide-polyvinylpyrrolidone hybrid films, we determine wavelength-dependent pseudo-optical constants and compression ratios that enable accurate prediction of optical modulation within the tested thickness range. Our methodology establishes a framework for engineering hybrid photochromic systems and demonstrates how data-driven modeling can overcome limitations in characterizing complex nanostructured materials.
\end{abstract}

\maketitle


\section{\label{sec:introduction} Introduction}

Photochromic films enable reversible optical modulation for smart windows, adaptive optics, and reconfigurable photonic devices \cite{deb1973optical,granqvist2000electrochromic}. Among the various stimuli that can drive optical switching in tungsten oxide systems, including electrochemical bias in electrochromic devices and thermal activation in thermochromic coatings, UV-driven photochromism is particularly attractive for passive, self-regulating glazing and wearable photonics because it requires no external circuitry or heat source \cite{wang2018advances, dong2022nanostructured}. Of the various photochromic architectures, hybrid systems comprising active nanoparticles dispersed within polymer matrices represent a particularly promising approach, exhibiting rapid photochromic response, high optical contrast, and ambient-condition reversibility \cite{kim2024amplifying,bourdin2019nanoparticles,kozlov2020photochromic,kim2024binder,pacheco2023tuning}. \cite{kim2024amplifying,bourdin2019nanoparticles,kozlov2020photochromic,kim2024binder,pacheco2023tuning}. Yet despite extensive material development \cite{wang2018advances,cong2016tungsten,bourdin2019nanoparticles,badour2023finetuned,bourdin2020coloring,chen2024dual,kim2024binder,zayim2025fabrication,kozlov2020photochromic,pacheco2023tuning,kim2024dispersibility,popov2020pvp}, rational design of these hybrid films remains hindered by a fundamental challenge: unlike homogeneous thin films with well-defined refractive indices and extinction coefficients, photochromic hybrid films possess no well-defined effective optical constants\cite{Polyanskiy2024}. Solution-processed deposition inevitably produces spatially inhomogeneous particle distributions, rendering the optical response dependent on fabrication-specific microstructure rather than tabulated material properties\cite{kim2024amplifying,kozlov2020photochromic,kim2024dispersibility}. This absence of predictive optical models forces iterative device development to rely on empirical trial-and-error.

Rigorous electromagnetic simulations offer one potential solution. Finite-difference time-domain (FDTD) and finite element methods can directly model particle distributions within films \cite{Oskooi2010,xue2023jax,Mahlau2026}, but their application to micron-scale photochromic layers encounters prohibitive computational cost and a deeper limitation: accurate predictions require precise knowledge of particle arrangements that vary unpredictably between fabrication runs. Effective medium theories provide computational efficiency by homogenizing optical properties through analytical mixing formulas, yet these approaches assume quasi-static conditions and spherical geometries violated in micron-scale films and fail to capture state-dependent photochromic transitions\cite{OzkanZayim2003, Zayim2005, Baydogan2007}. The predictive gap between assumed simulation geometries and realized experimental structures significantly constrains reliable design of photochromic devices\cite{Lam2025, MAHITHA2024237}.

Recent progress in photonic characterization demonstrates that optical properties can be extracted directly from experimental measurements rather than calculated from microscopic structure \cite{Palmer1985,Peng1994,Palmer1998,Sahoo2001,DeCrescent2016,Dutta2022,Alharshan2023,Bonal2021,Bass2025}. Data-driven approaches have successfully characterized homogeneous thin films through spectrophotometric measurements, including envelope methods, Kramers-Kronig analysis, and optimization-based techniques \cite{Palmer1985,Peng1994,Palmer1998,Alharshan2023,Bonal2021,Dutta2022}. Machine learning and deep learning methods have further advanced inverse characterization capabilities for photonic structures \cite{Unni2020,Chen2023,Ballester2024,KhairehWalieh2023,Pira2025}. These methods work well for uniform layers but do not address photochromic hybrid films, where the challenge is extracting effective state-dependent parameters that account for structural inhomogeneity across reversible optical transitions.

Here, we introduce a data-driven framework that extracts effective optical constants for photochromic hybrid films directly from minimal transmittance measurements. Our dual-state model approximates the inhomogeneous photochromic layer as a compressed homogeneous medium for each optical state (pristine and UV-irradiated). The model is characterized by wavelength-dependent pseudo-refractive indices and pseudo-extinction coefficients, along with compression factors that account for effective thickness reduction. These parameters are simultaneously optimized against experimental data from only a few samples, enabling accurate prediction of optical modulation across arbitrary film thicknesses. Implementation within a fully differentiable transfer matrix formulation \cite{danis2025tmmax} substantially reduces computational cost compared to full-wave simulations while maintaining physical interpretability \cite{Luce2022}. Validation with $\text{WO}_{3-x}$--PVP films \cite{kim2024amplifying,kozlov2020photochromic,popov2020pvp} demonstrates quantitative agreement with experimental spectra and successful interpolation to untested configurations, providing a pathway toward rational engineering of adaptive photonic devices.

\section{\label{sec:modeling} Data-Driven Effective Optical Modeling of Photochromic Micron-Scale Hybrid Films}

In this section, we formulate the numerical methodology employed for the optical modeling of photochromic micron-scale hybrid films. To elucidate the underlying principles of the model, we consider a generalized multilayer structure containing a single photochromically active layer. This active layer sits between sets of homogeneous layers, as defined by the following sequence:

\begin{equation}
S_{\text{P}} = \{M_{\text{inc}}, L_1^T, \dots, L_N^T, PC, L_1^B, \dots, L_M^B, M_{\text{sub}}\}
\label{eq:1}
\end{equation}

Here, $M_{\text{inc}}$ represents the incident medium, a semi-infinite layer where light enters the multilayer film, while $M_{\text{sub}}$ denotes the substrate medium where light exits the structure, also treated as semi-infinite because its thickness far exceeds that of the structure layers. $PC$ represents a hybrid layer consisting of photochromically active particles that possess an inhomogeneous distribution within a micron-scale matrix. The top layers $L_i^T$ ($i = 1, \dots, N$) and bottom layers $L_j^B$ ($j = 1, \dots, M$) are homogeneous layers with optical constants $n_{i,j}(\lambda)$, $k_{i,j}(\lambda)$, and thickness $d_{i,j}$. The homogeneous layers consist of materials with well-established optical constants documented in the literature\cite{Polyanskiy2024}. Both the refractive index and the extinction coefficient of these constituents are rigorously validated across a broad spectral range through extensive experimental benchmarks. In contrast, the hybrid photochromic ($PC$) film comprises nanometric active particles dispersed non-homogeneously, resulting in an optical response that is intrinsically dependent on the localized particle density within the matrix \cite{popov2020pvp, kozlov2020photochromic}. As a result, the hybrid film lacks a singular, well-defined set of intrinsic optical constants. While the optical response of passive layers is efficiently and accurately described using the Transfer Matrix Method (TMM)\cite{Abeles, macleod2010thin, katsidis2002general, harbecke1986coherent, byrnes2020multilayeropticalcalculations}, active hybrid layers cannot be treated with similar simplifications and instead require comprehensive numerical approaches such as 3D FDTD simulations \cite{Oskooi2010, Mahlau2026}. This requirement leads to a significant mismatch between the particle densities assumed in simulations and those realized experimentally, thereby complicating the design process and substantially increasing the computational cost\cite{taflove2005computational}. Consequently, this approach is impractical for systematic modeling and engineering of photochromic hybrid films.

\begin{figure}[ht!]
    \centering
    \includegraphics[width=\columnwidth]{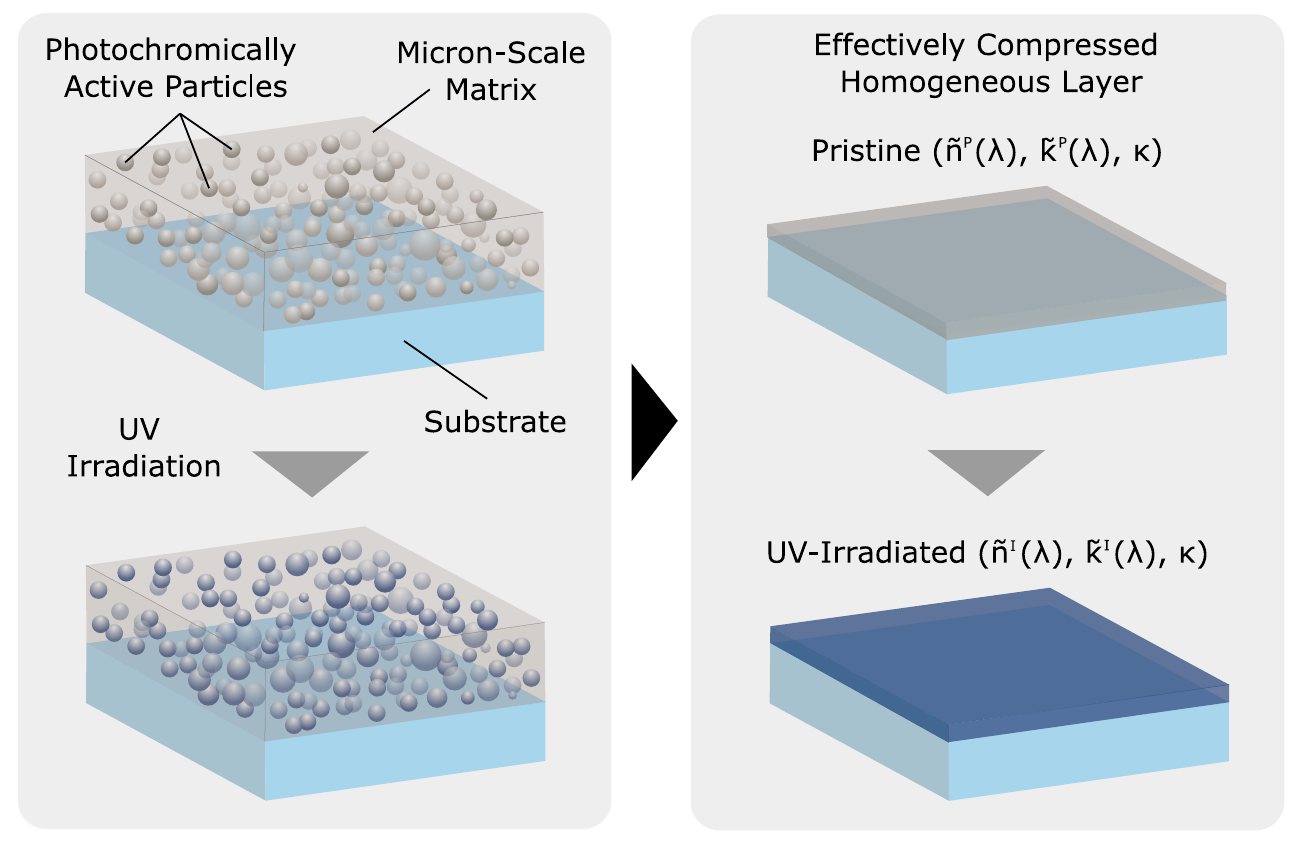}
    \caption{Effective compressed homogeneous layer approximation for photochromic hybrid films. The physical photochromic micron-scale hybrid film (left) consists of photochromically active particles dispersed non-homogeneously within a polymer matrix on a substrate. Upon UV irradiation, the particles undergo a photochromic transition, changing their optical properties. In our dual-state effective model (right), we approximate this complex inhomogeneous structure as an effectively compressed homogeneous layer for each state. The pristine state is characterized by pseudo-optical constants $\tilde{n}^P(\lambda)$, $\tilde{k}^P(\lambda)$ and compression factor $\kappa^P$, while the UV-irradiated state is described by $\tilde{n}^I(\lambda)$, $\tilde{k}^I(\lambda)$ and compression factor $\kappa^I$. This approximation enables efficient optical modeling using the transfer matrix method while capturing the essential optical response of the photochromic hybrid film through state-dependent effective optical properties.}
    \label{fig:figure1}
\end{figure}

\begin{figure*}[!ht]
    \centering
    \includegraphics[width=\textwidth]{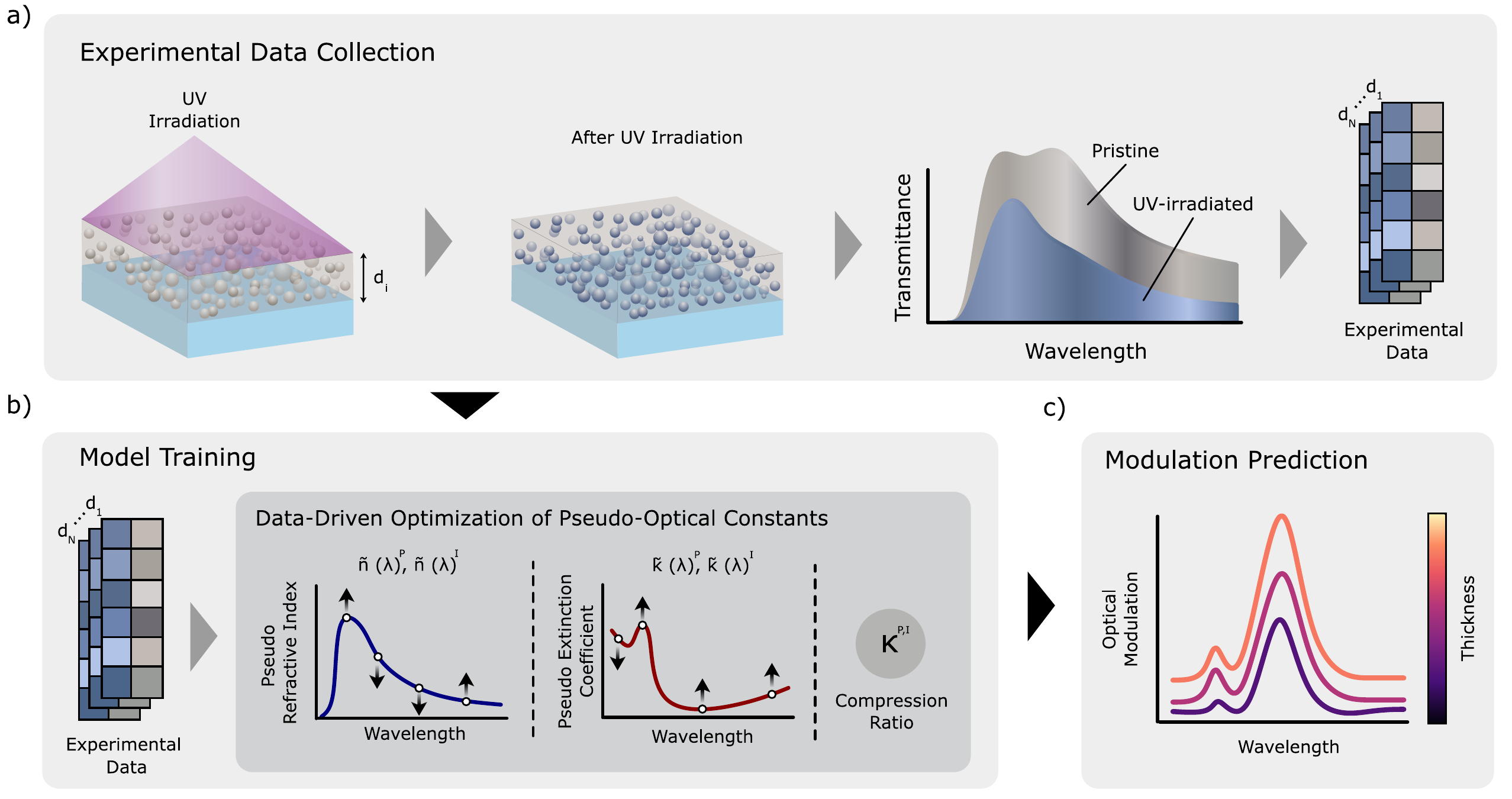}
    \caption{Data-driven optimization framework for dual-state effective modeling of photochromic micron-scale hybrid films. Pseudo-optical constants and compression factors are optimized through inverse engineering of experimental transmittance data, enabling optical modulation prediction across arbitrary film configurations. (a) Experimental data collection workflow: photochromic hybrid films of varying thickness are measured in both pristine and UV-irradiated states, yielding transmittance spectra pairs that constitute the training dataset. (b) Model training architecture: the transfer matrix method calculates wavelength-dependent transmittance using trainable pseudo-refractive indices ($\tilde{n}^P$, $\tilde{n}^I$), pseudo-extinction coefficients ($\tilde{k}^P$, $\tilde{k}^I$), and scalar compression ratios ($\kappa^P$, $\kappa^I$). These parameters are iteratively refined to minimize mean squared error between model predictions and experimental measurements across all samples, wavelengths, and incidence angles. (c) Predictive capability: once optimized, the model generates optical modulation spectra for film thicknesses beyond the training set, enabling rational design of photochromic devices with tailored switching characteristics.}
    \label{fig:figure2}
\end{figure*}

To overcome these limitations and simplify the modeling of photochromic hybrid systems, we introduce a dual-state effective model that categorizes the film based on its response to light exposure: the pristine (P) and UV-irradiated ($I$) states. These two states represent the physically well-defined endpoints of the photochromic switching of WO\textsubscript{3-x}-PVP hybrid films: the pristine state corresponds to the optically transparent, unactivated configuration achieved in the absence of any UV stimulus, while the UV-irradiated state corresponds to the colored, high-absorption configuration reached after saturation of the photochromic response under UV exposure. Both states are stable on experimental timescales and produce reproducible, thickness-dependent transmittance spectra, making them the natural training targets for a data-driven optical model. A single-state model would be insufficient to capture the optical modulation mechanism; the dual-state approach thus represents the minimum viable parameterisation that captures the full modulation range. In this model, we approximate the photochromic layer as an effectively compressed homogeneous medium for each respective state and define a refractive index and extinction coefficient for each case by assuming homogeneity, as illustrated in Fig.~\ref{fig:figure1}. Since these optical properties vary depending on the homogeneity of the photochromically active particles within the hybrid film, we define state-dependent pseudo-refractive indices and pseudo-extinction coefficients rather than unique optical constants. Furthermore, as each state is modeled as a compressed layer, a state-dependent compression factor is incorporated into the model. This factor represents the effective thickness of the micron-scale hybrid film when considered as a homogeneous layer. While this formulation features a single photochromic layer, the methodology is adaptable to structures with multiple photochromic layers. We determine the effective thickness for each state by multiplying the original thickness of the hybrid film by this factor. Under this pseudo layer approximation, the photochromic layer is mapped onto state-dependent pseudo layers. By treating the system as a state-dependent homogeneous layer with defined optical constants, we can employ the TMM \cite{danis2025tmmax} in the modeling process. This significantly accelerates the simulation procedure. Additionally, the pseudo-optical constants of the model are tuned through a data-driven training process. This approach substantially reduces the mismatch between experimentally observed optical responses and numerical predictions during the design process. Accordingly, the model's trainable parameter set,

\begin{equation}
\phi = \{\kappa^P, \kappa^I, \tilde{n}^P(\lambda), \tilde{n}^I(\lambda), \tilde{k}^P(\lambda), \tilde{k}^I(\lambda)\}
\label{eq:2}
\end{equation}

is optimized simultaneously. Here, the compression factors $\kappa^P$ and $\kappa^I$ are real scalars constrained to the interval $[0, 1]$ for the pristine and UV-irradiated states, respectively, while the pseudo-refractive indices $\tilde{n}^P(\lambda)$, $\tilde{n}^I(\lambda)$ and pseudo-extinction coefficients $\tilde{k}^P(\lambda)$, $\tilde{k}^I(\lambda)$ are wavelength-dependent vector quantities for each state.

We develop a training methodology to determine the optimal pseudo-optical constants and compression factors of the photochromic layer, as illustrated in Fig.~\ref{fig:figure2}. The parameter extraction problem is mathematically equivalent to nonlinear least-squares fitting: we seek the set $\phi$ that best reproduces the observed transmittance data, using iterative gradient descent rather than analytical inversion because the transfer matrix introduces a highly nonlinear dependence on $\phi$. To ensure close agreement between model predictions and experimental data, we conduct training on an experimental dataset comprising transmittance spectra pairs from several photochromic hybrid films (two samples in our implementation) with different thickness combinations, measured in both pristine and UV-irradiated states, as shown in Fig.~\ref{fig:figure2}(a). The experimental data collection requires only a few samples, as the pseudo-optical constants are optimized by fitting model predictions to experimental data, enabling accurate fits with minimal samples. Indeed, the model aims to generate reliable predictions across various thicknesses and/or material combinations using as few experimental data points as possible. The collected data subsequently serves as the basis for model training. During training, the model takes input parameters including thickness, material combination, incident angle, and wavelength, and produces transmission or reflection as output. As depicted in Fig.~\ref{fig:figure2}(b), the model performs transmission calculations at each wavelength, optimizing both the pseudo-refractive index and pseudo-extinction coefficient simultaneously. Each compression ratio is a scalar parameter that remains constant across all wavelengths. 

To generate model outputs during pseudo-optical constant optimization, we employ the TMM. The model computes the transmittance or reflectance of the photochromic micron-scale hybrid film by calculating the system matrix $M_{\text{total}}^{P,I}$ for each state (pristine or UV-irradiated):

\begin{equation}
\label{eq:total_matrix}
M_{\text{total}}^{P,I}
=
\prod_{i=1}^{N} M_{i}^{P,I}
\times
M_{\text{PC}}^{P,I}(\phi)
\times
\prod_{j=1}^{M} M_{j}^{P,I}
\end{equation}

This matrix calculation, commonly referred to as the Abeles TMM \cite{Abeles}, results from the successive multiplication of the transfer matrices of each layer and interface \cite{macleod2010thin, katsidis2002general, harbecke1986coherent, byrnes2020multilayeropticalcalculations}, including passive homogeneous dielectric layers and $M_{\text{PC}}^{P,I}$ in the middle, as expressed in Equation~(\ref{eq:total_matrix}).

\begin{equation}
M_i = I_i P_i = 
\begin{bmatrix}
\alpha_{i,i+1} & \gamma_{i,i+1} \\
\gamma_{i,i+1} & \alpha_{i,i+1}
\end{bmatrix}
\begin{bmatrix}
e^{-j\delta_i} & 0 \\
0 & e^{j\delta_i}
\end{bmatrix}
\label{eq:transfermatrix}
\end{equation}
Each layer is represented as $M_i = I_i P_i$, where $I_k$ denotes the interface matrix and $P_i$ represents the propagation matrix describing light propagation through the $i$-th layer, as expressed in Equation~(\ref{eq:transfermatrix}). The elements $\alpha_{i,i+1}$ and $\gamma_{i,i+1}$ vary depending on the polarization of the incoming light. The accumulated phase $\delta_i$ is given by:

\begin{equation}
\label{eq:interface_phase_terms}
\begin{aligned}
\alpha_{i,i+1} &=
\begin{cases}
\dfrac{ n_i\cos\theta_i + n_{i+1}\cos\theta_{i+1}}
      { 2 n_i\cos\theta_i } & \text{(s-pol.)} \\[6pt]
\dfrac{ n_i\cos\theta_{i+1} + n_{i+1}\cos\theta_i }
      { 2 n_i\cos\theta_i } & \text{(p-pol.)}
\end{cases} \\[12pt]
\gamma_{i,i+1} &=
\begin{cases}
\dfrac{ n_i\cos\theta_i - n_{i+1}\cos\theta_{i+1}}
      { 2 n_i\cos\theta_i} & \text{(s-pol.)} \\[6pt]
\dfrac{ n_i\cos\theta_{i+1} - n_{i+1}\cos\theta_i}
      { 2 n_i\cos\theta_i} & \text{(p-pol.)}
\end{cases} \\[12pt]
\end{aligned}
\end{equation}

The accumulated phase $\delta_i$ is given by:

\begin{equation}
\label{eq:delta}
\begin{aligned}
\delta_i &=
\begin{cases}
\dfrac{2\pi}{\lambda}\, n_i d_i \cos\theta_i 
& \text{(passive layers)} \\[6pt]
\dfrac{2\pi}{\lambda}\, n_i \kappa^{P,I} d_i  \cos\theta_i 
& \text{(photochromic layer)}
\end{cases}
\end{aligned}
\end{equation}

where complex-valued $n_i$ represents the sum of the refractive index and extinction coefficient (replaced by pseudo-refractive index $\tilde{n}^{P,I}$ and pseudo-extinction coefficient $\tilde{k}^{P,I}$ in photochromically active layers), $d_i$ denotes the layer thickness, $\theta_i$ represents the angle of incidence, and $\lambda$ denotes the wavelength of the incoming light. The elements of the resulting system matrix yield the reflection and transmission properties of the photochromic hybrid film in each state. The model performs this calculation at each wavelength and incident angle to predict the transmittance and reflectance of the hybrid film ($T_{\text{model}}^{P,I}(\lambda, \theta; \phi)$). We compute the mean square error against the available experimental data. We frame parameter extraction as a nonlinear inverse problem in which the model parameters $\phi$ are updated to minimize an objective function~\cite{Tarantola2005, Fujiwara2007, Dutta2022}, specifically the mean-squared error (MSE) between model-predicted and experimentally measured transmittances. The optimization objective is:

\begin{equation}
\label{eq:obj_function}
\mathcal{L}(\phi)
=
\sum_{\lambda_i \in \Lambda}
\sum_{\theta_i \in \Theta}
\left\|
T_{\text{model}}^{P,I}(\lambda_i,\theta_i;\phi)
-
T_{\text{exp}}^{P,I}(\lambda_i,\theta_i)
\right\|^{2}
\end{equation}

Here, $\Lambda$ and $\Theta$ denote the sets of sampled wavelengths and incidence angles, respectively. The squared norm provides a smooth, everywhere-differentiable objective compatible with gradient-based optimization. All parameters in $\phi$, including the wavelength-independent compression factors $\kappa^P$ and $\kappa^I$, contribute to the computed transmittance model through the accumulated phase term in Equation~(\ref{eq:delta}) and are therefore simultaneously updated during the minimization of $\mathcal{L}(\phi)$. The summation assigns equal weight to each wavelength sample, implicitly treating the spectrophotometer as having approximately uniform precision across the 200--1100 nm range. By optimizing the model with this objective function, we ensure that the pseudo-optical constants converge to best match the experimental data. We determine the parameters that minimize this objective. Once the optimal optical constants and compression ratios are identified, transmittance predictions can be generated for any desired thickness, wavelengths and incident angles. As shown in Fig.~\ref{fig:figure2}(c), predictions in each state naturally enable determination of optical modulation through this framework.

\section{\label{sec:filmpreparation}Photochromic Hybrid Film Preparation and Data Collection}

The training of our proposed model requires collecting experimental data from photochromic hybrid films. We select crystallized tungsten oxide ($\text{WO}_{3-x}$) particles dispersed within a polyvinylpyrrolidone (PVP) matrix as our photochromic hybrid system. Tungsten trioxide represents a highly versatile material whose high chemical stability and tunable color gradients enable precise tailoring of functional characteristics \cite{popov2020pvp, kozlov2020photochromic}. We collect experimental transmittance spectra pairs from several tungsten trioxide photochromic micron-scale hybrid films with different thickness combinations, measured in both pristine and UV-irradiated states.
\begin{figure}[ht!]
    \centering
    \includegraphics[width=\columnwidth]{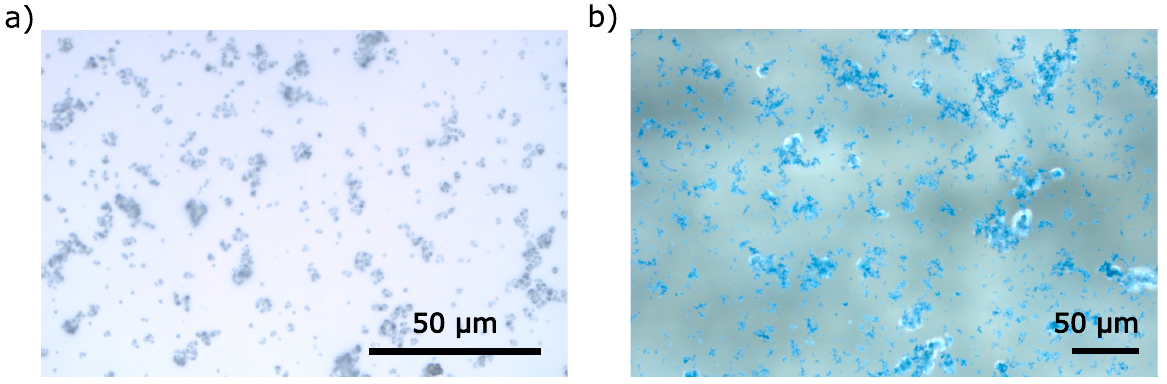}
    \caption{Microstructure of photochromic hybrid films. Optical microscope images of $\text{WO}_{3-x}$ hybrid photochromic single-layer film deposited at 1200 rpm, showing (a) pristine state and (b) UV-irradiated state. Scale bars, 50 $\mu\text{m}$. The images indicate that crystallized tungsten-oxide particles are embedded and non-uniformly distributed throughout the PVP matrix, highlighting the intrinsically inhomogeneous micro-scale morphology of the hybrid film.}
    \label{fig:figure3}
\end{figure}
Before coating, Corning glass substrates are cleaned with deionized water, followed by washing in an ultrasonic bath with ethanol. We use spin coating to coat the glass surface with the tungsten trioxide particle solution in the PVP matrix. The spin coating parameters, including rotation speed and deposition cycles, critically influence the structural and optical properties of $\text{WO}_{3-x}$ films, as the homogeneity of crystallized tungsten trioxide particles varies at different spin speeds \cite{brinker1990sol}. This behavior is showed in Fig.~\ref{fig:figure3}, where the microscopic analysis reveals the dispersion of $\text{WO}_{3-x}$ hybrid photochromic particles deposited via the single-layer coating method. The optical microscope image of the $\text{WO}_{3-x}$ hybrid photochromic single layer deposited at 1200 rpm reveals that crystallized tungsten oxide particles are dispersed within the PVP matrix. This indicates that the layer, which possesses a thickness on the order of microns, is not homogeneously composed of $\text{WO}_{3-x}$ material and exhibits a dispersed structure. Consequently, we perform coating in three distinct spin speed batches. Films are coated onto glass substrates using the spin coating method at 1200, 2000, and 2500 rpm for 60 seconds to both examine the system's performance across different uniformity conditions and investigate the pseudo-optical constants at these parameters. To preserve mechanical stability and achieve uniform films exceeding 150 $\mu$m in thickness, we apply multiple layers using the same configuration repeatedly. Each layer is carefully controlled to ensure uniformity and proper interlayer interaction. For model training, coatings are prepared at these three different spin speeds with two distinct thicknesses: single-layer and five-layer configurations. To validate that the model correctly maps the transmittance space after training, an intermediate thickness coating (three layers) is prepared at each of the three spin speeds, with the thickness falling within the range defined by the two training sample thicknesses.

The coated films reach steady state after 4 hours of equilibration at room temperature and atmospheric conditions, as they appear dark blue immediately after coating in their wet form and develop a light blue tinge upon drying. Upon UV irradiation, photons with energy exceeding the optical bandgap of WO$_{3-x}$ generate electron-hole pairs within the tungsten oxide crystallites \cite{wang2018advances, deb1973optical}. The photogenerated electrons partially reduce tungsten centres from W$^{6+}$ to W$^{5+}$, forming a tungsten bronze phase in conjunction with electron transfer from the surrounding PVP matrix, which acts as an electron donor \cite{kozlov2020photochromic}. These reduced tungsten sites support small polaron formation, giving rise to a broad intervalence charge-transfer absorption band centred in the near-infrared and extending into the visible, which is responsible for the characteristic blue coloration observed in Fig. 5 and the increase in pseudo-extinction coefficient beyond 500 nm visible in Fig. 4. It should be noted that, while bleaching of WO$_3$-based photochromic films via re-oxidation of W$^{5+}$ centres is documented in the literature, the strong coordination of PVP with WO$_3$ surface sites stabilises the reduced tungsten bronze state through charge-transfer interactions, resulting in slow and frequently incomplete bleaching in WO$_3$-PVP systems under ambient conditions \cite{Li2019ChemistrySelect, Wei2023Pacheco}. Accordingly, the UV-irradiated coloured state of the WO$_{3-x}$-PVP films studied here is effectively stable on the timescales of all transmittance measurements reported. The present work characterises the pristine and fully UV-coloured endpoint states; systematic cycling behaviour is beyond the scope of this study. The coatings are exposed to UV light for 10 minutes in an environment where the temperature does not exceed 30$^\circ$C. UV irradiation is carried out in a UV oven system (UV Crosslinker, UVP LLC) equipped with low-pressure mercury lamps emitting at 253.7 nm with an output power of 2.3 W. During these measurements, the laboratory temperature is maintained at 21$^\circ$C with a relative humidity of 42\%. Thin films coated on glass substrates are characterized for their photochromic properties by exposing them to UV light. The transmittance of these hybrid films in both their pristine and UV-irradiated states is measured over the 200--1100 nm spectral range using a Agilent 8453 UV-Visible spectrophotometer.

\section{\label{sec:results}Experimental Results and Model Predictions}

We apply our data-driven optimization framework to extract pseudo-optical constants from experimental transmittance measurements of $\text{WO}_{3-x}$--PVP hybrid films fabricated at three distinct spin-coating speeds (1200, 2000, and 2500 rpm). The training dataset comprises transmittance spectra from films with two thicknesses per spin speed (single-layer and five-layer configurations), measured in both pristine and UV-irradiated states. We first optimize the pseudo-optical constants and compression ratios, then validate the model against test transmittance spectra. Finally, we explore optical modulation behavior across a continuous thickness range (50--600 $\mu$m) to demonstrate the framework's predictive capabilities for photochromic hybrid films of arbitrary thickness.

\subsection{Optimization of Pseudo-Optical Parameters}

Through iterative optimization using the objective function defined in Equation~(\ref{eq:obj_function}), we determine the wavelength-dependent pseudo-refractive indices $\tilde{n}^P(\lambda)$ and $\tilde{n}^I(\lambda)$, pseudo-extinction coefficients $\tilde{k}^P(\lambda)$ and $\tilde{k}^I(\lambda)$, and compression factors $\kappa^P$, $\kappa^I$ that minimize the mean squared error between model predictions and experimental observations. We implement the model entirely within \texttt{katmer}, a high-performance, research-grade Python library specifically designed for the inverse characterization of optical multilayer thin films \cite{katmer2025}. Within \texttt{katmer}, we implement the formulation and numerical methodology described above in a fully differentiable manner, enabling gradient-based optimization \cite{deepmind2020jax} (Adam\cite{kingma2017adammethodstochasticoptimization}) for both hybrid micron-scale and homogeneous layers. We use the Adam optimizer with a learning rate of $10^{-3}$, first moment decay $\beta_1 = 0.9$, second moment decay $\beta_2 = 0.999$, and numerical stability constant $\epsilon = 10^{-8}$. All models are trained for 150 iterations. For all three spin-speed conditions, the pseudo-optical constants and compression factors are initialized with identical starting values: $n(\lambda) = 1.5$ for all wavelengths, $k(\lambda) = 1.0$ for all wavelengths, and $\kappa = 1 \times 10^{-3}$. This common initialization ensures that differences between the extracted pseudo-optical constants across spin-speed conditions arise solely from the experimental transmittance data rather than from initialization bias. The \texttt{katmer} library builds upon two core packages: \texttt{tmmax} [42], a JAX-based \cite{jax2018github} TMM engine that leverages just-in-time compilation and vectorized operations \cite{xla2023github} for rapid evaluation of optical properties across complex multilayer stacks, and Equinox \cite{kidger2021equinox}, a lightweight neural network library with a pytree-based architecture that provides the flexibility required for constructing sophisticated models while ensuring efficient parameter management. Further code details are available in our GitHub repository.

\begin{figure}[ht!]
\centering
\includegraphics[width=\columnwidth]{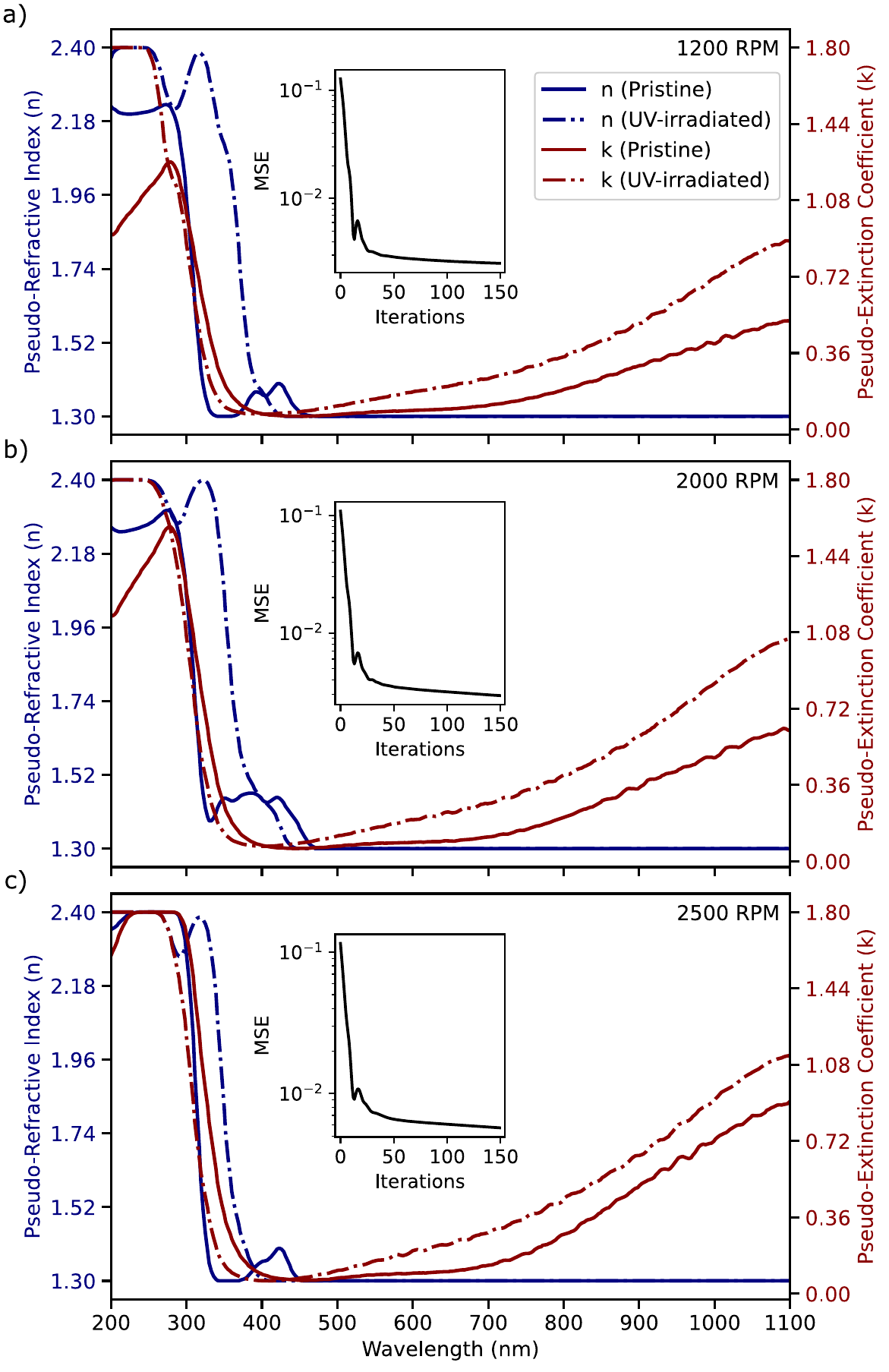}
\caption{Optimized pseudo-optical constants for photochromic hybrid films. Wavelength-dependent pseudo-refractive indices (blue) and pseudo-extinction coefficients (red) extracted through inverse engineering for films deposited at (a) 1200 rpm, (b) 2000 rpm, and (c) 2500 rpm. Solid lines represent pristine state parameters ($\tilde{n}^P$, $\tilde{k}^P$), while dashed lines indicate UV-irradiated state parameters ($\tilde{n}^I$, $\tilde{k}^I$). Insets show mean squared error versus training iteration (150 iterations), demonstrating convergence of the optimization process.}
\label{fig:figure4}
\end{figure}

Our spectrophotometer measures transmittance at normal incidence across 901 sample points spanning 200--1100 nm. We select 350 wavelength points for $\tilde{n}$ and $\tilde{k}$, as the pseudo-optical constants do not exhibit rapid, physically abrupt variations and thus do not require parameters at every measured wavelength. Since we measure transmittance in both pristine and UV-irradiated states and include two compression ratio parameters, each spin speed yields a total of 702 tunable parameters. In our multilayer simulations, the top and bottom layers are air and glass, respectively, with optical constants taken from established literature without modification \cite{Peck:72, Rodriguez-deMarcos:16}. Also, model predictions are generated across all 901 wavelength points. During model training, we impose upper and lower boundaries on both pseudo-refractive index and pseudo-extinction coefficient to prevent physically unrealistic predictions. We establish these boundaries manually, setting the pseudo-refractive index range to 1.3--2.4 and the pseudo-extinction coefficient range to 0--1.8. These boundaries apply identically to both pristine and UV-irradiated states and yield the lowest objective values in our tests. Minor boundary adjustments do not substantially alter the extracted pseudo-optical constant profiles. In Figs.~\ref{fig:figure4}(a)--(c), the pristine-state pseudo-refractive index (solid blue lines) ranges from 2.2--2.4 near 200 nm, decreases toward 300 nm, exhibits a peak near 400 nm, then remains constant around 1.3 from 500 to 1100 nm. In the UV-irradiated state (dashed blue lines), the pseudo-refractive index peaks near 200 nm and 350 nm ($\approx 2.4$), then decreases and stabilizes at 1.3 beyond 350 nm. The key difference between states is a 50 nm shift: the pristine state decreases rapidly after 300 nm, while the UV-irradiated state decreases after 350 nm. The pseudo-extinction coefficients reveal distinct absorption features that capture the optical modulation mechanism. For 1200 rpm films (Fig.~\ref{fig:figure4}(a)), the pristine-state pseudo-extinction coefficient (solid red line) begins near 1.8 at 200 nm, peaks around 300 nm, then decreases to nearly zero by 400 nm. For 2000 rpm (Fig.~\ref{fig:figure4}(b)), it starts at 2.0, and for 2500 rpm (Fig.~\ref{fig:figure4}(c)), at 1.75, each peaking at 300 nm before decreasing to near-zero absorption at 400 nm. In the UV-irradiated state (dashed red lines), all three spin speeds show the pseudo-extinction coefficient beginning at 1.75 near 200 nm and gradually decreasing toward 400 nm. Beyond 400 nm, both states exhibit a slow increase in pseudo-extinction coefficient toward 1100 nm. At 1100 nm, the pristine-state values are approximately 0.5 (1200 rpm), 0.6 (2000 rpm), and 0.9 (2500 rpm), while the UV-irradiated state values reach approximately 0.9 (1200 rpm), 1.0 (2000 rpm), and 1.1 (2500 rpm). The variation in pseudo-optical constant behavior across spin speeds arises from the increased coating homogeneity at higher spin speeds. Macroscopically, photochromic hybrid layers with different homogeneities exhibit distinct optical responses, explaining the spin-speed dependence of these pseudo-refractive indices. The compression ratio varies across both states and spin speeds. For 1200 rpm, $\kappa$ is $5.7 \times 10^{-4}$ in the pristine state and $6.5 \times 10^{-4}$ in the UV-irradiated state. For 2000 rpm, these values are $6.7 \times 10^{-4}$ and $7.8 \times 10^{-4}$, respectively. For 2500 rpm, they are $7.1 \times 10^{-4}$ and $8.8 \times 10^{-4}$. When micron-scale thicknesses (50--600 $\mu$m) are compressed using these coefficients, the effective thicknesses become several hundred nanometers, justifying our use of coherent TMM and confirming the consistency of our simulations.
\begin{figure*}[ht!]
\centering
\includegraphics[width=\textwidth]{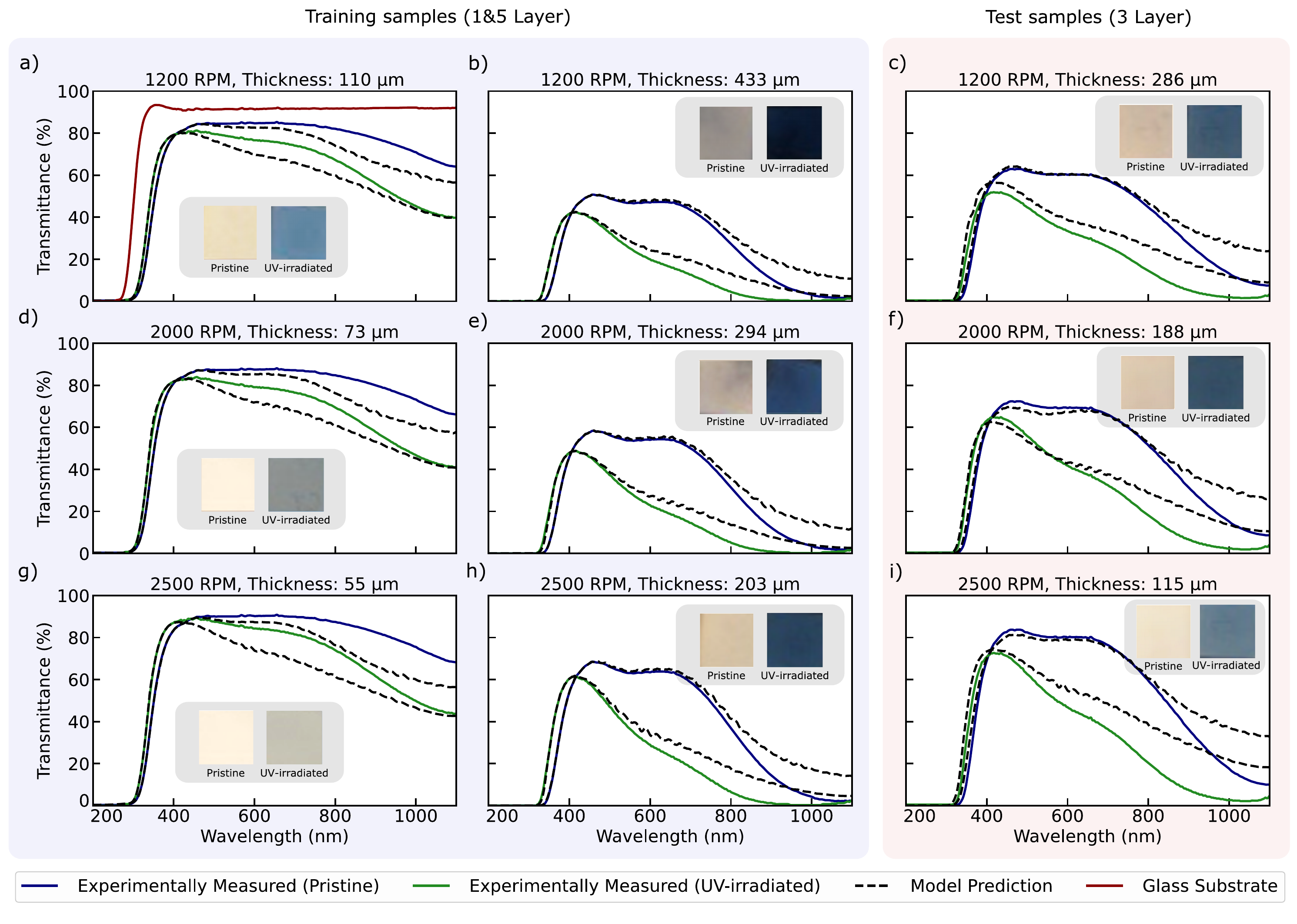}
\caption{Experimental transmittance and transmission predictions of the dual-state effective model. Measured transmittance spectra (solid lines: blue for pristine state, green for UV-irradiated state) compared against model predictions (dashed black lines) for $\text{WO}_{3-x}$--PVP hybrid films with varying thicknesses. Red lines indicate bare glass substrate transmittance. Subplots (a--c) show 1200 rpm films, (d--f) show 2000 rpm films, and (g--i) show 2500 rpm films. Training samples include single-layer (a, d, g) and five-layer (b, e, h) configurations, while intermediate three-layer films (c, f, i) serve as test cases. Photographic insets display the visible color change of the $\text{WO}_{3-x}$--PVP hybrid films upon UV irradiation. The model accurately reproduces experimental measurements across all configurations, demonstrating reliable thickness interpolation and state-dependent optical response prediction.}
\label{fig:figure5}
\end{figure*}
The compression ratio increases with spin speed for both states, consistent with enhanced homogeneity and reduced agglomeration of $\text{WO}_{3-x}$ crystalline particles at higher speeds, allowing the films to be better approximated as coherent homogeneous layers. We train all models for 150 iterations across all states and spin speeds. In all three cases, the mean squared error decreases and stabilizes by 150 iterations, as shown in the insets of Fig.~\ref{fig:figure4}. For 1200 rpm and 2000 rpm (Figs.~\ref{fig:figure4}(a)--(b), insets), the MSE decreases from $10^{-1}$ to $3\times10^{-3}$, evaluated over 901 spectral points on the [0, 1] transmittance scale, corresponds to an RMS transmittance deviation of approximately 5.5 percentage points across the full 200--1100 nm range, which is within the expected variability of spin-coated film thickness uniformity. For 2500 rpm (Fig.~\ref{fig:figure4}(c), inset), a converged MSE of $6\times10^{-3}$ correspondingly yields an RMS deviation of approximately 7.7 percentage points.

A transient non-monotonic feature is visible in all three curves at approximately iteration 20. This behavior is attributable to the early-iteration dynamics of the Adam optimizer: the bias-corrected effective learning rate is not monotonic in early steps due to the different decay rates of the first moment ($\beta_1=0.9$) and second moment ($\beta_2=0.999$) estimates, which can produce temporary parameter overshoots before the adaptive step size stabilizes \cite{Kingma2015}. Identical initial parameter values across all three spin-speed optimizations produce nearly the same early-iteration trajectory, placing the bump at the same iteration index in all three insets. The loss landscape defined by the 702 coupled parameters is inherently non-convex through the transfer matrix formulation, and transient non-monotonic behavior in early training is therefore expected and characteristic of gradient-based optimization in high-dimensional non-convex settings \cite{Dauphin2014}.

\subsection{Experimental vs Model-Predicted Transmittance Spectra}

To validate the accuracy of the optimized pseudo-optical constants, we compare model predictions against experimental transmittance measurements for both training samples and previously unseen test films with intermediate thicknesses. Fig.~\ref{fig:figure5} presents comprehensive transmittance spectra for films prepared at the three spin-coating speeds. Experimental measurements for pristine films (blue solid lines) and UV-irradiated films (green solid lines) are overlaid with corresponding model predictions (black dashed lines) calculated using the optimized pseudo-optical parameters and compression ratios from Fig.~\ref{fig:figure4}. The bare glass substrate transmittance (red line) serves as the reference baseline, exhibiting nearly complete transparency beyond 300 nm. In Fig.~\ref{fig:figure5}, each row corresponds to one coating speed, with the first two columns showing transmittance data used during model training. The first column represents thin single-layer films, while the second column represents thick five-layer films. The third column shows test films with intermediate thicknesses not included in training, validating the model's interpolation capability. For all training samples (Figs.~\ref{fig:figure5}a--b, \ref{fig:figure5}d--e, \ref{fig:figure5}g--h), the model fits the experimental data accurately. For test films (Figs.~\ref{fig:figure5}c, \ref{fig:figure5}f, \ref{fig:figure5}i), the model demonstrates excellent predictive capability across all spin-coating configurations. This demonstrates the framework's predictive power: once pseudo-optical constants are extracted from minimal experimental data, the model reliably interpolates optical responses for arbitrary thicknesses within the explored range. Examining predictions with optimized optical constants across all training transmittance spectra reveals negligible transmittance differences from 200--600 nm, with minor deviations from 600--1100 nm. Minor deviations between model predictions and experimental measurements in the 600--1100 nm range reflect the spectral resolution of the measurement system, while the model maintains physical consistency and wavelength correlation throughout the entire spectral range.

The photographic insets in each panel visually confirm the photochromic transition, showing the characteristic color change from pale yellow-brown (pristine) to dark blue (UV-irradiated) that accompanies the measured transmittance modulation. The systematic thickness dependence observed across panels reveals the design space accessible through this modeling approach: thinner films maintain higher visible transmittance while sacrificing modulation depth, whereas thicker films achieve stronger optical contrast at the expense of reduced overall transmission. This quantitative prediction of the transmittance--thickness trade-off enables rational selection of film parameters to meet specific application requirements.

\subsection{Optical Modulation Predictions}

Having validated model accuracy against both test and training transmission spectra at different spin speeds, we leverage the framework to predict optical modulation characteristics across a continuous range of film thicknesses and wavelengths without requiring additional experiments. Optical modulation $\Delta T(\lambda, d)$ is defined as the difference between pristine and UV-irradiated state transmittances: $\Delta T(\lambda, d) = |T^P(\lambda, d) - T^I(\lambda, d)|$. Using the optimized optical constants and compression ratios, we examine how optical modulation varies continuously across thicknesses from 50 to 600 $\mu$m for photochromic hybrid $\text{WO}_{3-x}$--PVP films.

\begin{figure}[ht!]
\centering
\includegraphics[width=\columnwidth]{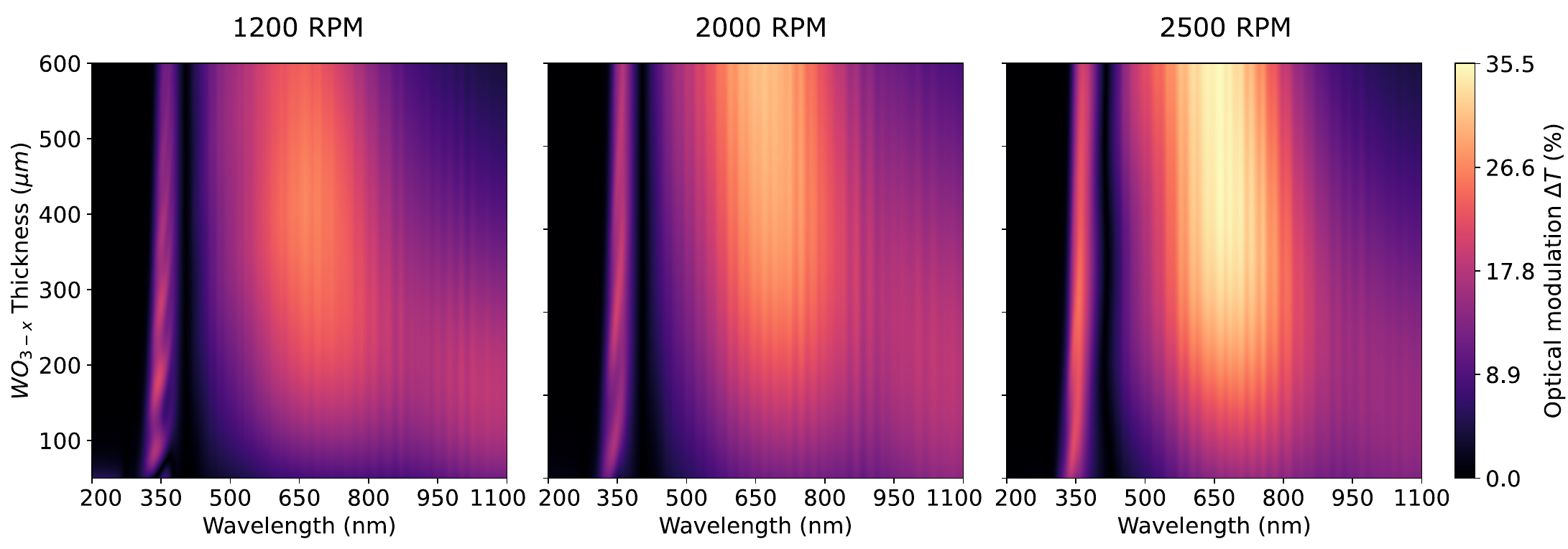}
\caption{Predicted optical modulation distribution across thickness and wavelength. Two-dimensional maps showing optical modulation $\Delta T = |T^P - T^I|$ calculated using optimized pseudo-optical constants for films deposited at (a) 1200 rpm, (b) 2000 rpm, and (c) 2500 rpm. The color scale indicates modulation magnitude in percentage points. The maps reveal thickness-dependent optimization regimes, with maximum modulation occurring at intermediate thicknesses (200--400 $\mu$m) and wavelengths in the visible to near-infrared range (500--800 nm). These comprehensive predictions enable rational film design for target applications.}
\label{fig:figure6}
\end{figure}

Fig.~\ref{fig:figure6} presents two-dimensional modulation maps generated by the model for each spin-coating condition, spanning thicknesses from 50 to 600 $\mu$m and wavelengths from 200 to 1100 nm. These maps reveal the rich design space accessible through thickness engineering of photochromic hybrid films. The modulation exhibits strong wavelength dependence, with peak values occurring in the visible and near-infrared regions where the pseudo-refractive index and pseudo-extinction coefficient change most dramatically between pristine and UV-irradiated states (as seen in Fig.~\ref{fig:figure4}). All three modulation predictions exhibit two distinct peaks. A smaller modulation peak appears near 350 nm, arising from pseudo-refractive index differences between states. A broader, larger peak (500--850 nm) centers near 650 nm, originating from the increasing divergence in pseudo-extinction coefficients between pristine and UV-irradiated states beyond 500 nm (see Fig.~\ref{fig:figure4}), which continues to 1100 nm. For 1200 rpm films (Fig.~\ref{fig:figure6}(a)), maximum modulation approaches 27\% in the 500--650 nm range for intermediate thicknesses around 200--400 $\mu$m. Films deposited at 2000 rpm (Fig.~\ref{fig:figure6}(b)) achieve slightly higher peak modulation ($\approx 32\%$), while 2500 rpm films (Fig.~\ref{fig:figure6}(c)) reach maximum values exceeding 35\%. This enhancement with increasing spin speed arises from improved coating homogeneity. The thickness-dependent behavior shows that very thin films ($<100$ $\mu$m) provide insufficient absorption change to generate substantial modulation. Across all configurations, increased spin-coating speed yields enhanced modulation performance.

These predictive maps enable application-specific design optimization without requiring exhaustive experimental fabrication and characterization. For example, smart window applications prioritizing visible light modulation while maintaining high luminous transmittance would favor thicknesses of 150--250 $\mu$m at 2000 rpm, whereas near-infrared switching applications might favor thicker films at 2500 rpm to maximize modulation beyond 800 nm. The ability to generate such comprehensive design guidelines from only six training measurements (two thicknesses at three spin speeds) demonstrates the power of the data-driven optical constant extraction approach for photochromic hybrid film engineering.

Although the modulation maps in Fig.~\ref{fig:figure6} display predictions over a continuous thickness range from 50 to 600 $\mu$m, the conditions governing the model's predictive validity beyond the training configurations are physically well-defined and derive directly from the structure of the framework. The pseudo-optical constants and compression ratios are trained as thickness-independent quantities; combined with a coherent transfer matrix formalism that is analytic in accumulated phase, this means that transmittance predictions for thicknesses outside the training set follow from the same governing equations rather than from curve fitting, making the model's reach naturally broader than its training configurations alone. At the lower end, the fabrication process defines the physically relevant minimum: the thinnest training sample at each spin speed corresponds to the minimum film thickness achievable at that rotation rate, and there is therefore no meaningful target for prediction below this value within the spin-coating framework used here. At the upper end, the binding constraint arises from the coherency assumption embedded in the model. Because the photochromic layer is treated as a coherent homogeneous medium with effective thickness $k^{P,I} d$, this representation is physically valid only as long as the effective thickness remains within the coherent interference regime, where phase relationships between partial waves are preserved. Once the effective thickness exceeds this limit, partially coherent or incoherent transfer matrix formulations are required \cite{katsidis2002general, harbecke1986coherent, Santbergen2013, Centurioni2005, byrnes2020multilayeropticalcalculations}, and the coherent limit on effective thickness therefore sets the practical upper bound on the range of physical film thicknesses over which the framework can be reliably applied.

\section{\label{sec:conclusion}Conclusion}

We introduce a data-driven framework that overcomes fundamental limitations in the optical modeling of photochromic micron-scale hybrid films by extracting effective optical constants directly from minimal experimental measurements. Our dual-state effective model approximates inhomogeneous photochromic layers as compressed homogeneous media characterized by wavelength-dependent pseudo-refractive indices and pseudo-extinction coefficients, enabling accurate prediction of optical responses across a wide range of film thicknesses while substantially reducing computational cost compared to full-wave electromagnetic simulations. Through vectorization and optimization within the transfer matrix formalism, transmittance calculations are reduced to seconds. Validation with $\text{WO}_{3-x}$--PVP hybrid films demonstrates that the framework achieves quantitative agreement with experimental transmittance spectra using a limited number of training samples, successfully interpolating optical behavior for untested configurations and generating comprehensive design maps that guide thickness optimization for target applications. The methodology is readily applicable to diverse photochromic and stimuli-responsive material systems, offering a practical pathway from empirical trial-and-error toward rational engineering of adaptive photonic devices for smart windows and reconfigurable optics.

\section*{Acknowledgments}

This work was supported by the Istanbul Technical University Scientific Research Projects Coordination Unit under project no.: MGA-2017-40594, and by the Scientific and Technological Research Council of Türkiye (TUBITAK) under the 2209-A Research Project Support Programme for Undergraduate Students, 2022 First-Term Call. Financial support was provided by Latvian Council of Science Project No. lzp-2024/1-0632. Also, SZK acknowledge SWEB project 101087367 funded by the HORIZON-WIDERA-2022-TALENTS-01-01.

\section*{Data Availability}

All relevant data generated or analyzed in this study can be obtained from the corresponding author on a reasonable request.

\section*{Code Availability}

The model training and analysis code in this study is publicly available at \url{https://github.com/bahremsd/inv-eng-photochromic-optics}. The repository includes Jupyter notebooks that implement the optimization using the \texttt{katmer} library in Python and can be run to reproduce all results presented in this work.

\def\bibsection{\section*{References}}
\bibliographystyle{apsrev4-2}  
\bibliography{main}
\end{document}